\documentclass[]{article}
\usepackage{color}
\usepackage{amsmath,amsfonts,amssymb}
\usepackage{graphicx}
\usepackage{jabbrv}
\usepackage{authblk}
\usepackage{url}
\usepackage{textcomp}
\usepackage{caption}
\usepackage{appendix}
\usepackage{hyperref}
\usepackage{gensymb}
\hypersetup{hidelinks}
\usepackage{geometry}
\usepackage{xcolor}

\usepackage{subcaption}
\usepackage{lineno}
\usepackage{gensymb}
\usepackage{marvosym}
\usepackage{makecell}
\title{Chip-integrated extended-cavity mode-locked laser in the visible}

\author[1, 2, *, \Cross]{Lisa V. Winkler}
\author[1, \Cross]{Govert Neijts}
\author[1]{Hubertus J. M. Bastiaens}
\author[3]{Melissa J. Goodwin}
\author[1, 6]{Albert van Rees}
\author[4]{Philip P. J. Schrinner}
\author[4]{Marcel Hoekman}
\author[4]{Ronald Dekker}
\author[5]{Adriano R. do Nascimento Jr.}
\author[1]{Peter J. M. van der Slot}
\author[2]{Christian Nölleke}
\author[1]{Klaus-J. Boller}

\affil[1]{Laser Physics and Nonlinear Optics, Faculty of Science and Technology, MESA+ Institute, University of Twente, Enschede, the Netherlands}
\affil[2]{TOPTICA Photonics AG, Gräfelfing, Germany}
\affil[3]{Faculty of Electrical Engineering, Mathematics and Computer Science (EEMCS), Nanolab, University of Twente, Enschede, the Netherlands}
\affil[4]{LioniX International B.V., Enschede, The Netherlands}
\affil[5]{PHIX B.V., Enschede, the Netherlands}
\affil[6]{Currently at Chilas B.V}
\affil[{\Cross}]{These authors contributed equally to this work}

\affil[*]{l.v.winkler@utwente.nl}

\begin{document}

\maketitle

\begin{abstract}
Mode-locked lasers are of interest for applications such as biological imaging, non-linear frequency conversion, and single-photon generation. In the infrared, chip-integrated mode-locked lasers have been demonstrated through integration of laser diodes with low-loss photonic circuits. However additional challenges, such as a higher propagation loss and smaller alignment tolerances have prevented the realization of such lasers in the visible range.
Here, we demonstrate the first chip-integrated mode-locked diode laser in the visible using an integrated photonic circuit for cavity extension. Based on a gallium arsenide gain chip and a low-loss silicon nitride feedback circuit, the laser is passively mode-locked using a saturable absorber implemented by focused ion beam milling.
At a center wavelength of 642~nm, the laser shows an average output power of 3.4\,mW, with a spectral bandwidth of 1.5\,nm at a repetition rate of 7.84\,GHz.

\end{abstract}

\section{Introduction}
Photonic chip integration of mode-locked laser sources operating in the visible spectral range holds significant promise for various applications, including Raman scattering in biological samples for chemical imaging \cite{ozeki2012high}, on-chip nonlinear frequency conversion \cite{Jankowski:20} to extend the range of integrated light sources, or the excitation of single photon emitters, contributing to scalable quantum processing with integrated photonics \cite{IntegratedSPE10.1116/5.0011316}.

Many semiconductor mode-locked lasers operating in the infrared have been demonstrated, with an overview given in \cite{hermans_-chip_2022}. However, there is only a single report of an integrated mode-locked diode laser in the visible spectral range \cite{vasilev_mode_2013}, which utilizes a monolithic InGaN cavity with no external feedback.
Such monolithic mode-locked semiconductor lasers typically suffer from limited coherence due to high intrinsic losses in the semiconductor amplifiers. Furthermore, the short cavity lengths lead to high pulse repetition rates. 

These limitations can be overcome by integration of semiconductor laser diodes with photonic chips comprising low-loss integrated waveguide circuits. Increasing the photon lifetime in the cavity leads to a lower optical and radio frequency (RF) linewidth \cite{paschotta2004noise, paschotta_optical_2006}, while the cavity length can be chosen to match a desired repetition rate. In recent years, this approach has successfully been demonstrated for infra-red lasers, e.g. in \cite{Vissers:21, wang2017iii}. However, no such lasers have been demonstrated in the visible.
This is due to additional, intrinsic challenges posed by shorter wavelengths. These are in particular a higher propagation loss in photonic chips due to increased scattering, smaller alignment tolerances, and a lack of suitable integrated saturable absorbers. 

Here, we present the first hybrid-integrated mode-locked laser in the visible spectral range, opening a new spectral window for integrated mode-locked lasers. Our laser operates at a center wavelength of 642~nm, with an average output power of 3.2\,mW and a spectral bandwidth of 1.5\,nm at a repetition rate of 7.84\,GHz.

\section{Design and Fabrication}
\begin{figure}[ht]
\centering
\includegraphics[width=\linewidth]{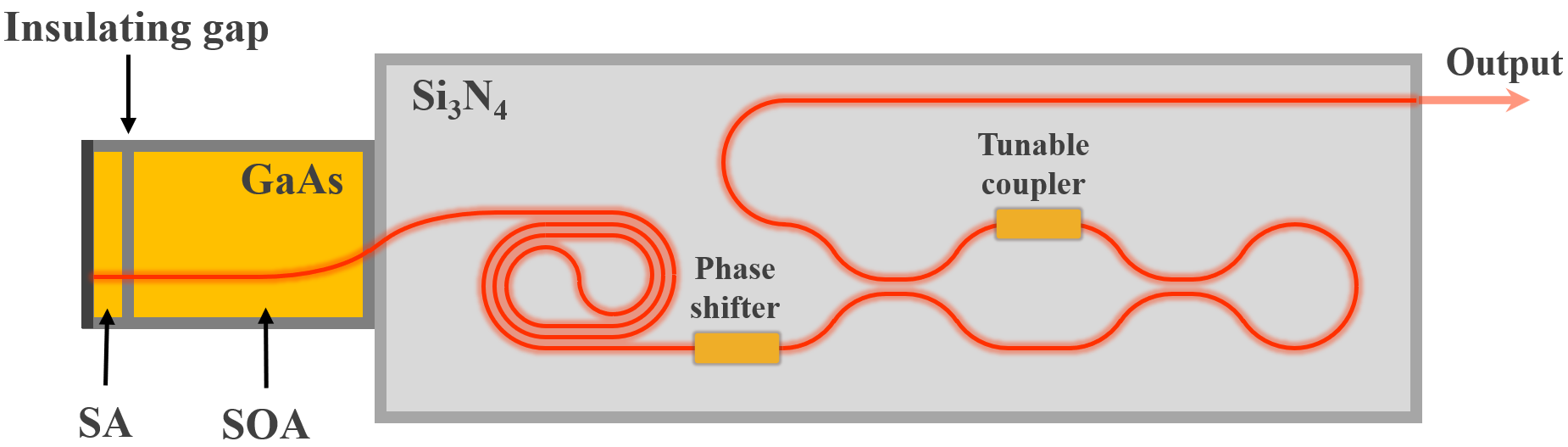}
\caption{Schematic of the hybrid integrated semiconductor laser cavity. Left in yellow illustrates the gain chip, comprising a semiconductor optical amplifier (SOA) and saturable absorber (SA) section. Right in grey shows the layout of the passive feedback chip, with a waveguide spiral, phase shifter, and tunable coupler together with a Sagnac loop, serving as a tunable mirror.
All waveguides are shown in red.}
\label{fig:schematic}
\end{figure}

The laser consists of a gallium-arsenide semiconductor optical amplifier (SOA), for the generation and amplification of light, and a feedback waveguide chip for extending the cavity length as shown in Fig. \ref{fig:schematic}. The feedback chip is fixed on a submount and packaged with an output fiber and electrical connections.  The two chips are edge-coupled, to form the hybrid laser.

The feedback chip comprises silicon-nitride (Si$_3$N$_4$) waveguides with an asymmetric double-stripe cross-section, offering low propagation loss (0.32$\pm$0.05\,dB/cm) for wavelengths at around 642\,nm, as presented in  \cite{Winkler:24}. The waveguide circuit includes an integrated Sagnac loop mirror, which functions as one of the cavity mirrors. This mirror provides tunable outcoupling by varying the electrical power supplied to a resistive heater at one of the arms of the tunable coupler. Another resistive heater is placed on top of part of the waveguide and acts as a phase shifter, enabling tuning of the longitudinal cavity modes. A waveguide spiral provides a passive extension of the cavity, extending the photon lifetime. The geometrical path length of the passive photonic integrated circuit is about 10.8\,mm, corresponding to an optical single-pass path length of 16.3\,mm, using a calculated effective refractive index of 1.51. On the other side of the chip, a lateral taper facilitates mode matching to the SOA. The waveguides of the SOA and the feedback chip are angled with respect to the facet to prevent back reflections in the guided fundamental mode.

The gain chip (EXALOS AG, center wavelength: 642\,nm) comprises a 0.80\,mm long gallium-arsenide gain section with a highly reflective coating ($\ge$ 95\%) on the back facet. The gain chip adds 2.6\,mm to the single-pass optical cavity length, based on an effective refractive index of 3.2. When the gain chip is edge-coupled to the feedback chip, they form a hybrid laser cavity with a total single-pass optical path length of 18.9\,mm, corresponding to a calculated free spectral range of 7.9\,GHz. At a wavelength of 642\,nm, this translates to a mode separation of approximately 10\,pm, and a roundtrip time of about 126\,ps.

To achieve passive mode-locking of the laser, we use a saturable absorber (SA). For semiconductor lasers, SAs can be made from the same material as the SOA, with the SOA operating in forward bias, and the SA operating in reverse bias. The reverse bias of the SA shortens the recovery time of the absorber, allowing for increased absorption when the SA is not saturated by the high-intensity light of pulses \cite{coldren_diode_2012, karin_ultrafast_1994}.

We use a commercially available gain section, which is originally designed and manufactured as a single continuous structure. For the required independent biasing, we separated the gain section into two electrically isolated parts: one serving as the SOA and the other as the SA. 
To achieve this separation, we create an insulating gap in the gain chip's top electrode using focused ion beam (FIB) milling \cite{giannuzzi_review_1999}, a technique that has previously been applied to implement SAs in infrared, monolithic, modelocked lasers \cite{barbarin:hal-02549373}. This approach allows for quick prototyping using readily available gain sections.

\begin{figure}[ht]
\centering
\includegraphics[width=\linewidth]{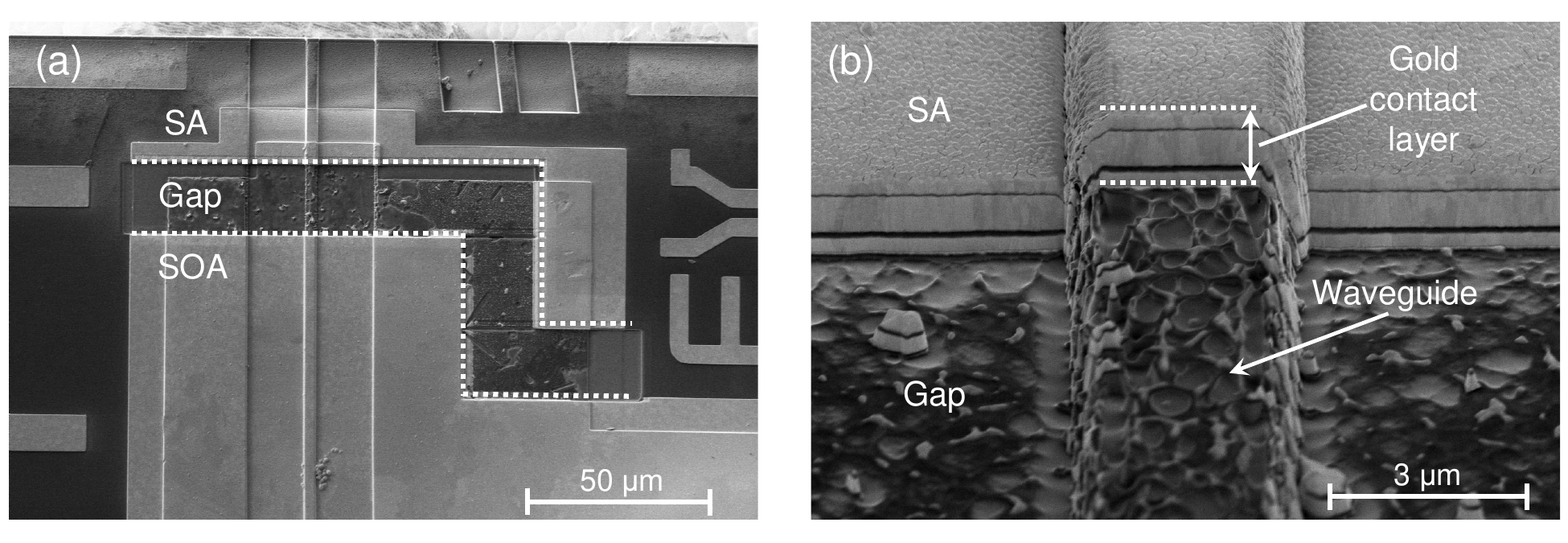}
\caption{SEM image of a cut made using FIB milling, illustrating the insulating gap where the conducting top layer was removed, separating the SA (above) from the SOA (below). (b) Zoomed-in SEM image at the edge of the SA, showing the removal of the gold contact layer and exposed ridge waveguide.}
\label{fig:SEM-images}
\end{figure}

FIB milling is a high-precision micro-fabrication technique that involves a finely focused beam of ions incident onto a conducting substrate. The primary mechanism at work is physical sputtering, where the momentum of the ions in the beam displaces atoms from the surface of the substrate, thereby removing material in a highly controlled manner. This allows for sub-micrometer precision in material removal \cite{giannuzzi_review_1999}. We target the conductive layer on top of the gallium-arsenide substrate with a beam of gallium ions, at a beam current of 2.6\,nA, to selectively address only that layer and not the SOA's ridge waveguide.  A scanning electron microscope (SEM) image (Fig. \ref{fig:SEM-images}a) shows the final cut, illustrating the SA and SOA sections, separated by the insulating gap. A zoomed-in SEM image of the border between the gap and the SA (Fig. \ref{fig:SEM-images}b) provides a detailed view of the gap, showing that the gold (light tint) is removed from the top of the ridge waveguide (dark tint) in the gap.
The insulating gap has a length of approximately 20\,$\mu$m, with a resulting SA length of approximately 17.5\,$\mu$m. This value is in the typical 2 to 4\%-range of SA length with regard to the SOA length (0.8\,mm) reported for optimal mode-locking performance \cite{javaloyes_mode-locking_2010, barbarin:hal-02549373}.
The electrical isolation achieved with this approach is measured to be 10\,k$\Omega$, indicating successful separation of the SA from the SOA.

\section{Results}

\begin{figure}[ht]
\centering
\includegraphics[width=\linewidth]{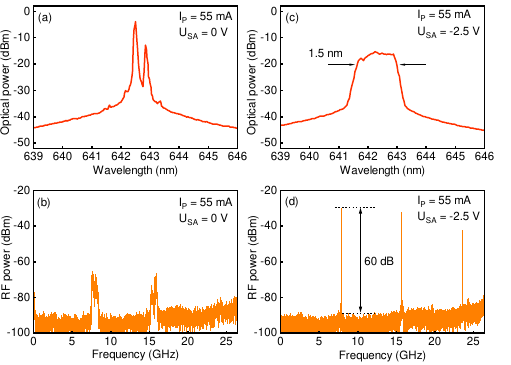}
\caption{(a) Optical spectrum of hybrid laser in multi-mode operation, at 55~mA pump current, and without bias applied to SA. (b) Corresponding RF spectrum of the laser in multi-mode operation. (c) Optical spectrum of the laser operated in a mode-locked state, with a reverse bias of -2.5~V, showing the 1.5~nm broad spectrum, resembling a frequency comb. (d) Corresponding RF spectrum of the mode-locked laser, showing the strong and sharp RF beat notes at integer multiples of the cavity FSR.}
\label{fig:hero-measurement}
\end{figure}

To investigate the initiation of mode-locking via the SA, we characterize the laser output in the optical and RF domains for two SA bias conditions (0\,V and -2.5\,V), which enables clear discrimination between multi-mode and mode-locked operation (Fig.~\ref{fig:hero-measurement}). For these measurements, the outcoupling ratio is fexed to 75\% and the laser pump current is fixed to the maximum safe operating current of 55\,mA. For both SA settings, we record the optical spectrum using an optical spectrum analyzer (OSA, Ando AQ-6315A) with the resolution set to 0.1\,nm.  As this resolution is insufficient for resolving the individual longitudinal modes, the OSA measurement only indicates the range of longitudinal modes at which the laser oscillates. For further increasing the resolution, RF spectra are simultaneously recorded with a fast photodiode (Thorlabs DXM12CF, 12~GHz bandwidth) and an electrical spectrum analyzer (ESA, Keysight CXA Signal Analyzer N9000B, 26.5-GHz bandwidth), with a resolution bandwidth (RBW) of 3\,MHz and a video bandwidth (VBW) of 300\,kHz. 

At a reverse bias voltage of 0\,V, the laser operates with an average output power of 3.7\,mW. The optical spectrum (a) displays a typical multi-mode profile, with several distinct peaks, centered around a wavelength of 642\,nm. The corresponding RF spectrum (b) shows two broad, irregular, and unstable peaks, near the free spectral range (FSR) of the laser cavity, with a width of around 1~GHz at the -10-dB level. A second set of scattered peaks is visible at around the second harmonic of the FSR.
These spectral features imply that multiple longitudinal, non-equidistant modes are oscillating in the laser cavity, indicating the absence of mode-locking. Hence, without reverse-biasing of the SA, the laser exhibits multi-mode operation.

When applying a reverse bias of -2.5 V to the SA, the increased intra-cavity losses decrease average output power only slightly (3.4 mW). However, the laser output spectra (c) change dramatically. The optical spectrum now shows a broadened flat-top profile without any specifically preferred wavelengths, which is characteristic for mode-locked lasers. The change indicates the redistribution of energy across the spectrum. The spectrum spans a full-width half-maximum (FWHM), of approximately 1.5\,nm at the same center wavelength of 642\,nm. The measured RF spectra show a most dramatic change. In contrast to operation with no SA bias, the RF spectrum now shows sharp, strong RF beat notes at the cavity FSR of 7.84\,GHz as well as the second and third harmonic, with an amplitude of 60\,dB above the noise floor of the ESA. These sharp RF beat notes are a clear sign of mode-locking with equidistant light frequencies, also called optical comb source. With a 1.5\,nm spectral bandwidth, and a mode separation of about 10\,pm at 642\,nm, around 150 different modes are expected to oscillate simultaneously in this mode-locked state of the hybrid laser. 

The passive stability of the repetiton rate of a mode-locked laser can be quantified by recording the main line shape components of the laser's RF spectrum. A Voigt fit of the fundamental beat note, recorded with a RBW of 300~Hz, yields an RF linewidth of 40\,kHz (FWHM), with a Lorentzian component of 11\,kHz, and a Gaussian component of 33\,kHz. This low linewidth, approximately five orders of magnitude smaller than the repetition rate, indicates stable mode-locking, which compares well with other mode-locked lasers extended to similar cavity lengths, typically operating at 1550\,nm \cite{hermans_-chip_2022, davenport_integrated_2018, Vissers:21}.

\begin{figure}[ht]
\centering
\includegraphics[width=\linewidth]{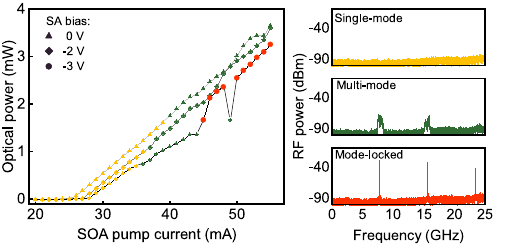}
\caption{(a) Optical output power of the hybrid laser, as a function of the SOA pump current, for different reverse bias voltages. (b) Typical observed RF spectra of different operational states of the laser. No RF peaks illustrate single-mode, broad unstable peaks multi-mode, and sharp and strong peaks mode-locked operation}
\label{fig:PI-and-regimes}
\end{figure}

To investigate whether mode locking can also be achieved at other operating parameters, we measure the average laser output power as a function of pump current for three different values of reverse bias applied to the SA section (Fig. \ref{fig:PI-and-regimes}a): 0\,V, -2\,V, and -3\,V. Simultaneously, RF spectra are recorded with a fast photodiode and the ESA. Depending on the pump current and SA bias, the RF spectra show three different operating regimes, with typical examples shown in Fig. \ref{fig:PI-and-regimes}b): single-mode oscillation, identified by the absence of RF peaks (yellow trace), multi-mode oscillation indicated by broad and unstable peaks in the RF signal (green trace), and mode-locked operation, marked by sharp and stable RF beat notes (red trace). For a bias voltage of 0\,V, we observe the lowest threshold at a pump current of 26\,mA and a slope efficiency of 141$\pm$2\,mW/A. With increasing reverse bias, the threshold increases due to increased absorption in the SA. For pump currents just above threshold, the laser always emits a single longitudinal mode, as indicated with yellow data symbols. At somewhat higher pump currents of about 35 to 40\,mA multi-mode operation of the laser sets in, indicated with green symbols.

For a reverse bias of 0~V and -2~V, the laser exhibits multi-mode operation up to the maximum applicable pump current of 55\,mA. However, for a reverse bias voltage of -3~V, the laser enters a mode-locked state at a pump current of 44~mA, characterized by sharp RF beat notes and indicated by red symbols. The mode-locked state is lost, and changes back to multi-mode, at a pump current of 49\,mA. At 50\,mA, the laser enters a stable mode-locked regime, which is maintained up to the maximum investigated pump current of 55\,mA. From this, we conclude that for a reverse bias voltage of -3\,V, the saturation levels and the recovery times of the SA and of the SOA balance each other, resulting in stable mode-locking, for a broad range of pump currents. The maximum average output power of the laser in this mode-locked state, at a pump current of 55\,mA, is approximately 3.2\,mW, which is slightly lower than then average output power measured at -2.5~V due to the increased SA absorption.

\section{Conclusion}
To conclude, we have demonstrated the first hybrid integrated, mode-locked semiconductor laser operating in the visible spectral range. This is achieved through the hybrid integration of commercially available semiconductor laser diodes and low-loss Si$_3$N$_4$ integrated photonic feedback circuits. Using FIB milling to electrically isolate part of the SOA section to function as an SA is a versatile approach for exploring mode-locking in a wide range of wavelengths. As a next step, gain chips with optimized SA and SOA parameters could also be manufactured lithographically. With diode amplifiers readily available across a wide range of visible wavelengths, from near UV to near IR, and the continued maturation of the Si$_3$N$_4$ photonic circuit platform, it seems feasible to realize chip-integrated ultrafast devices for a wide range of visible wavelengths, as has recently been demonstrated for continuous sources \cite{corato-zanarella_widely_2023}.
With the scalability provided by the photonic integrated platforms and modern semiconductor technology, visible wavelength hybrid-integrated mode-locked semiconductor lasers have the potential to contribute applications as portable biological imaging \cite{shapiro_effects_2011}, optical communications \cite{oubei_48_2015}, and portable atomic clocks \cite{beard_two-photon_2024, sharma_analysis_2022}.

\subsection*{Acknowledgments}
We thank Yvan Klaver, Akhileshwar Mishra, Anzal Memon and Thomas Puppe for fruitful discussions and the Nonlinear Nanophotonics group for providing measurement equipment.

\subsection*{Disclosures}
The authors declare no conflicts of interest.

\section*{Funding}
Rijksdienst voor Ondernemend Nederland (PPS\_2020\_90)

\bibliographystyle{opticajnl}
\bibliography{sample}

\begin{thebibliography}{10}
\newcommand{\enquote}[1]{``#1''}

\bibitem{ozeki2012high}
Y.~Ozeki, W.~Umemura, Y.~Otsuka, \emph{et~al.}, \enquote{High-speed molecular spectral imaging of tissue with stimulated raman scattering,} {\protect\JournalTitle{Nature Photonics}} \textbf{6}, 845--851 (2012).

\bibitem{Jankowski:20}
M.~Jankowski, C.~Langrock, B.~Desiatov, \emph{et~al.}, \enquote{Ultrabroadband nonlinear optics in nanophotonic periodically poled lithium niobate waveguides,} {\protect\JournalTitle{Optica}} \textbf{7}, 40--46 (2020).

\bibitem{IntegratedSPE10.1116/5.0011316}
J.~Lee, V.~Leong, D.~Kalashnikov, \emph{et~al.}, \enquote{{Integrated single photon emitters},} {\protect\JournalTitle{AVS Quantum Science}} \textbf{2}, 031701 (2020).

\bibitem{hermans_-chip_2022}
A.~Hermans, K.~Van~Gasse, and B.~Kuyken, \enquote{On-chip optical comb sources,} {\protect\JournalTitle{APL Photonics}} \textbf{7}, 100901 (2022).

\bibitem{vasilev_mode_2013}
P.~P. Vasil'ev, A.~B. Sergeev, I.~V. Smetanin, \emph{et~al.}, \enquote{Mode locking in monolithic two-section {InGaN} blue-violet semiconductor lasers,} {\protect\JournalTitle{Applied Physics Letters}} \textbf{102}, 121115 (2013).

\bibitem{paschotta2004noise}
R.~Paschotta, \enquote{Noise of mode-locked lasers (part ii): timing jitter and other fluctuations,} {\protect\JournalTitle{Applied Physics B}} \textbf{79}, 163--173 (2004).

\bibitem{paschotta_optical_2006}
R.~Paschotta, A.~Schlatter, S.~C. Zeller, \emph{et~al.}, \enquote{Optical phase noise and carrier-envelope offset noise of mode-locked lasers,} {\protect\JournalTitle{Applied Physics B: Lasers and Optics}} \textbf{82}, 265--273 (2006).

\bibitem{Vissers:21}
E.~Vissers, S.~Poelman, C.~O. de~Beeck, \emph{et~al.}, \enquote{Hybrid integrated mode-locked laser diodes with a silicon nitride extended cavity,} {\protect\JournalTitle{Opt. Express}} \textbf{29}, 15013--15022 (2021).

\bibitem{wang2017iii}
Z.~Wang, K.~Van~Gasse, V.~Moskalenko, \emph{et~al.}, \enquote{A iii-v-on-si ultra-dense comb laser,} {\protect\JournalTitle{Light: Science \& Applications}} \textbf{6}, e16260--e16260 (2017).

\bibitem{Winkler:24}
L.~V. Winkler, K.~Gerritsma, A.~van Rees, \emph{et~al.}, \enquote{Widely tunable and narrow-linewidth hybrid-integrated diode laser at 637 nm,} {\protect\JournalTitle{Opt. Express}} \textbf{32}, 29710--29720 (2024).

\bibitem{coldren_diode_2012}
L.~A. Coldren, S.~W. Corzine, and M.~Mashanovitch, \emph{Diode lasers and photonic integrated circuits}, no. 218 in Wiley series in microwave and optical engineering (Wiley, Hoboken, N.J, 2012), 2nd ed.

\bibitem{karin_ultrafast_1994}
J.~R. Karin, R.~J. Helkey, D.~J. Derickson, \emph{et~al.}, \enquote{Ultrafast dynamics in field-enhanced saturable absorbers,} {\protect\JournalTitle{Applied Physics Letters}} \textbf{64}, 676--678 (1994).

\bibitem{giannuzzi_review_1999}
L.~Giannuzzi and F.~Stevie, \enquote{A review of focused ion beam milling techniques for {TEM} specimen preparation,} {\protect\JournalTitle{Micron}} \textbf{30}, 197--204 (1999).

\bibitem{barbarin:hal-02549373}
Y.~Barbarin, E.~Bente, M.~J.~R. Heck, \emph{et~al.}, \enquote{{18 GHz Fabry-P{\'e}rot integrated extended cavity passively modelocked lasers},} in \emph{{European Conference on Integrated Optics (ECIO)},}  (Copenhagen, Denmark, 2007).

\bibitem{javaloyes_mode-locking_2010}
J.~Javaloyes and S.~Balle, \enquote{Mode-{Locking} in {Semiconductor} {Fabry}-{Pérot} {Lasers},} {\protect\JournalTitle{IEEE Journal of Quantum Electronics}} \textbf{46}, 1023--1030 (2010).

\bibitem{davenport_integrated_2018}
M.~L. Davenport, S.~Liu, and J.~E. Bowers, \enquote{Integrated heterogeneous silicon/{III}–{V} mode-locked lasers,} {\protect\JournalTitle{Photonics Research}} \textbf{6}, 468 (2018).

\bibitem{corato-zanarella_widely_2023}
M.~Corato-Zanarella, A.~Gil-Molina, X.~Ji, \emph{et~al.}, \enquote{Widely tunable and narrow-linewidth chip-scale lasers from near-ultraviolet to near-infrared wavelengths,} {\protect\JournalTitle{Nature Photonics}} \textbf{17}, 157--164 (2023).

\bibitem{shapiro_effects_2011}
M.~J. Shapiro, C.~C. Chow, P.~A. Karth, \emph{et~al.}, \enquote{Effects of {Green} {Diode} {Laser} in the {Treatment} of {Pediatric} {Coats} {Disease},} {\protect\JournalTitle{American Journal of Ophthalmology}} \textbf{151}, 725--731.e2 (2011).

\bibitem{oubei_48_2015}
H.~M. Oubei, J.~R. Duran, B.~Janjua, \emph{et~al.}, \enquote{48 {Gbit}/s 16-{QAM}-{OFDM} transmission based on compact 450-nm laser for underwater wireless optical communication,} {\protect\JournalTitle{Optics Express}} \textbf{23}, 23302 (2015).

\bibitem{beard_two-photon_2024}
R.~Beard, K.~W. Martin, J.~D. Elgin, \emph{et~al.}, \enquote{Two-photon rubidium clock detecting 776 nm fluorescence,} {\protect\JournalTitle{Optics Express}} \textbf{32}, 7417 (2024).

\bibitem{sharma_analysis_2022}
A.~Sharma, S.~Kolkowitz, and M.~Saffman, \enquote{Analysis of a {Cesium} lattice optical clock,}  (2022). ArXiv:2203.08708 [physics, physics:quant-ph].

\end{thebibliography}

\end{document}